\documentstyle[12pt,epsf]{article}
\newcommand{\HH}{{\cal H}}

\newcommand{\wt}{\widetilde}
\newcommand{\wh}{\widehat}

\newcommand{\bd}{\bar{\rm D}}

\newcommand{\pip}{\phi^{i\prime}}

\newcommand{\be}{\begin{equation}}
\newcommand{\ee}{\end{equation}}
\newcommand{\ben}{\begin{eqnarray}\displaystyle}
\newcommand{\een}{\end{eqnarray}}
\newcommand{\refb}[1]{(\ref{#1})}
\newcommand{\p}{\partial}
\newcommand{\sectiono}[1]{\section{#1}\setcounter{equation}{0}}

\begin{document}

{}~ \hfill\vbox{\hbox{hep-th/0003124}\hbox{MRI-PHY/P20000306}
}\break

\vskip 3.5cm

\centerline{\large \bf Vortex Pair Creation on Brane-Antibrane Pair}
\medskip
\centerline{\large \bf  via Marginal Deformation}

\vspace*{6.0ex}

\centerline{\large \rm Jaydeep Majumder\footnote{E-mail:
joydeep@mri.ernet.in} and Ashoke Sen
\footnote{E-mail: sen@mri.ernet.in}}

\vspace*{1.5ex}

\centerline{\large \it Mehta Research Institute of Mathematics}
 \centerline{\large \it and Mathematical Physics}

\centerline{\large \it  Chhatnag Road, Jhoosi,
Allahabad 211019, INDIA}

\vspace*{4.5ex}

\centerline {\bf Abstract}

It has been conjectured that the vortex solution on a D-brane -
anti-D-brane system represents a D-brane of two lower dimension. We
establish this result by first identifying a series of marginal
deformations which create the vortex - antivortex pair on the brane -
antibrane system, and then showing that under this series of marginal
deformations the original D-brane - anti-D-brane system becomes a D-brane
- anti-D-brane system with two lower dimensions. Generalization of this
construction to the case of solitons of higher codimension is also
discussed. 

\vfill \eject

\tableofcontents

\baselineskip=18pt

\sectiono{Introduction and Summary} \label{s1}

BPS
D-brane - anti-D-brane system of type IIA and IIB string theories admit
tachyonic modes\cite{9511194,9604091}. It has been conjectured that at the
minimum of the
tachyon potential the tension of the brane-antibrane system is exactly
canceled by the negative value of the tachyon potential, so that at this
point the brane-antibrane system is indistinguishible from the
vacuum\cite{9805019,9805170}.  It
has been further conjectured that various solitonic solutions on the
brane-antibrane pair, where the tachyon approaches the minimum of the
potential asymptotically, represent various lower dimensional branes.
Thus for example, on a single D$p$-$\bd p$ brane system, the kink solution
represents a non-BPS D-$(p-1)$
brane\cite{9806155,9808141,9809111,9812031,9812135} of type II string
theory,
whereas a vortex solution represents a BPS D-$(p-2)$ brane of the same
theory\cite{9808141,9810188}. There are generalizations of this conjecture
in which
solitons of higher codimension on more than one pair of D-brane -
$\bd$-brane system correspond to BPS and non-BPS D-branes of codimension
$>2$\cite{9808141,9810188,9812135}.\footnote{Recently evidence for some of
these
conjectures, and similar
conjectures\cite{RECK,9902105} involving D-branes in
bosonic
string theories, have been
found\cite{9912249,0001084,0002117,0002211,0002237,0003031} 
using string field
theory\cite{WITTENBSFT,WITTENSFT}. This approach uses the level truncation
scheme developed by
Kostelecky and
Samuel\cite{KS,AREF}. These conjectures have also been analysed
using
renormalization group flow on the world-sheet theory\cite{0003101,0003110}
following earlier work of ref.\cite{9406125}.}

The codimension one case, {\it i.e.} the identification of the kink
solution on the D$p$-brane $\bd p$-brane pair with a non-BPS
D-$(p-1)$-brane, has been demonstrated explicitly (although indirectly) by
identifying a series of marginal deformations which interpolate between
the D-brane - $\bd$-brane pair and the kink solution, and showing that
this series of deformations take the original D$p$-brane $\bd p$-brane
system to a non-BPS D-$(p-1)$-brane\cite{9808141,9903123}. This is done by
compactifying one of the directions tangential to the brane-antibrane
system on a circle, switching on half unit of Wilson line on the
antibrane, and reducing the radius of the circle to a critical radius
where the lowest mode of the tachyon becomes marginal. One then uses this
marginal deformation to create the kink and then takes the radius of the
circle back to infinity.

It is natural to ask if this procedure can be generalised to the case of
vortex and higher codimension solitons on the brane-antibrane pair. One
faces the following problem for the vortex. The original
D$p$-brane $\bd p$-brane
system is neutral under Ramond-Ramond (RR) gauge fields since the RR
charge of the brane and the anti-brane cancel. But a vortex, being
identified to a BPS D-$(p-2)$ brane, carries RR charge. Since RR charge is
quantized, one cannot hope to have a continuous interpolation between
these two configurations.\footnote{It has been suggested by Maldacena that
by switching on background magnetic field as in
ref.\cite{9503014}
one may be able to resolve this problem\cite{MALPRI}. The approach
followed in \cite{9703217,9704006} might also be helpful.} What one can
hope
to do however is to find a marginal deformation which interpolates between
the original  D$p$-brane $\bd p$-brane system and a vortex - antivortex
pair on this system. If we
can show that this marginal deformation converts the boundary conformal
field theory (BCFT) associated with the D$p$-brane $\bd p$-brane system to
the BCFT associated with the D-$(p-2)$-brane $\bd$-$(p-2)$-brane system,
then we would establish the equivalence between a vortex solution and a
D-$(p-2)$-brane.

This is the problem we address in this paper. The steps involved in the
analysis, which have been summarised in section \ref{s2}, are more or
less the same as the ones used for showing the equivalence of the kink
solution on the brane-antibrane pair with a codimension one non-BPS brane. 
For convenience of notation we study the case of a vortex solution on a
D2-brane $\bd$2-brane system. We compactify both directions tangential to
the brane, switch on appropriate Wilson lines, and reduce the radii of the
torus to certain critical values where the tachyonic deformation
corresponding to the creation of the vortex-antivortex pair becomes a
marginal deformation. We then switch on this marginal deformation and
study the fate of the BCFT under this marginal deformation. These steps
have been discussed in detail in section \ref{s3}. In section \ref{s4} we
study the effect of increasing the radii of the compact directions back to
large values. We show that under this series of deformations the original
BCFT gets deformed to a new BCFT describing the dynamics of open strings
on a D0-$\bd$0 pair. This establishes the equivalence of a vortex solution
on a D2-brane $\bd$2-brane pair, and a D0-brane. In section \ref{s5} we
discuss generalization of this analysis to solutions of higher
codimension. 

\sectiono{The General Strategy} \label{s2}

In this section we shall outline the general strategy that we shall
follow in order to establish the equivalence between a vortex-antivortex
pair on a BPS D$p$-brane - $\bd p$ brane system, and a D-$(p-2)$ -
$\bd$-$(p-2)$ brane pair. For definiteness we
shall take our starting point to be a 2-brane - $\bar 2$-brane pair of
type IIA string theory, but the analysis can clearly be generalised to any
$p$. 

The steps are as follows: 
\begin{enumerate}
\item We take a parallel 2-brane - $\bar 2$-brane pair of type IIA string
theory along $x^1-x^2$ plane, and compactify the $x^1$ and $x^2$
directions on circles of radii $R_1$ and $R_2$ respectively.
There are four different Chan-Paton (CP) sectors of open string 
states. States with CP factors proportional to the $2\times 2$ identity
matrix $I$ and the Pauli matrix $\sigma_3$, representing
open string states with both ends on the same brane, have the
standard GSO projection under which the Neveu-Schwarz (NS) sector ground
state is taken to be odd. On the other hand states with CP factors
$\sigma_1$ and $\sigma_2$, representing open strings with two ends on two
different branes, have opposite GSO projection, and in particular contain
tachyonic modes. We shall identify the tachyonic modes associated with the
sectors $\sigma_1$ and $\sigma_2$ with the real and the imaginary parts
respectively of a complex tachyon field $T$.
\item We switch on half a unit of Wilson line along each of these circles.
This makes open strings with CP factors $\sigma_1$ and $\sigma_2$,
including the tachyon field $T$, antiperiodic along each of the two
circles. Thus we can expand the tachyon field as
\be \label{e2.1}
T(x^1, x^2, t) = \sum_{m,n\in Z} T_{m+{1\over 2}, n+{1\over 2}}(t) e^{i (m
+{1\over 2}) {x^1\over R_1} + i(n+{1\over 2}) {x^2\over R_2}}\, .
\ee
The mass of the mode $T_{m+{1\over 2}, n + {1\over 2}}$ is given by
\be \label{e2.2}
(M_{m+{1\over 2}, n+{1\over 2}})^2 = {(m+{1\over 2})^2\over (R_1)^2}
+  {(n+{1\over 2})^2\over (R_2)^2} -{1\over 2}\, ,
\ee
in the $\alpha'=1$ unit that we shall be using.
\item {}From eq.\refb{e2.2} we see that for $(R_1)^{-2}+(R_2)^{-2}>2$
there are no
tachyonic modes. For $(R_1)^{-2}+(R_2)^{-2}=2$ the modes $T_{\pm{1\over
2},\pm{1\over 2}}$ become marginal. We shall see in section \ref{s3} that
at $R_1=R_2=1$ the deformation corresponding to 
\ben \label{e2.3}
T(x^1, x^2) &=& -i\alpha [e^{{i\over 2}(x^1+x^2)} - e^{-{i\over
2}(x^1+x^2)}
+ i(e^{{i\over 2}(x^1-x^2)} - e^{-{i\over 2}(x^1-x^2)})] \nonumber \\
&=& 2 \alpha (\sin{x^1+x^2\over 2} + i \sin{x^1 - x^2\over
2})\, ,
\een
for arbitrary constant $\alpha$
becomes exactly marginal, {\it i.e.} switching on this vev of the tachyon
does not cost any energy. 
$T(x^1,x^2)$ has zeroes at $x^1=x^2=0$ and at $x^1=x^2=\pi$. Near
$x^1=x^2=0$,
\be \label{e2.5}
T\simeq (1+i) \alpha (x^1 - ix^2)\, .
\ee
Thus it looks like a vortex.\footnote{Of course whether we call this a
vortex or an anti-vortex is a matter of convention. Once we fix the
convention for the vortex, then a configuration with opposite orientation
can be identified as an anti-vortex.} On the other hand, near
$x^1=x^2=\pi$,
\be \label{e2.6} 
T\simeq (-1+i) \alpha ((x^1-\pi) + i (x^2 -\pi))\, .
\ee
This looks like an anti-vortex. Thus we see that  switching on a tachyon
background of the form \refb{e2.3} corresponds to creating a
vortex-antivortex pair.
\item After switching on the deformation \refb{e2.3} we take the radii
$R_1$ and $R_2$ back to infinity, since we want to describe a
vortex-antivortex pair on infinite D2-brane $\bd$2-brane system. Here we
find that for a generic value of $\alpha$, once we increase $R_1$, $R_2$
beyond 1, there is a one point function of the modes $T_{\pm{1\over 2},
\pm{1\over 2}}$. This is not surprising, since for $R_i>1$, the
deformation \refb{e2.3} is a relevant perturbation, and hence
switching on
this background breaks conformal invariance of the world-sheet theory.
However, we find that besides the trivial background $\alpha=0$, there is
another inequivalent point (which we shall choose to be $\alpha=1$ by
suitably normalizing $\alpha$) where the one point function of
$T_{\pm{1\over 2},\pm{1\over 2}}$ vanish, and hence the theory is
conformally invariant. We shall identify the $\alpha=1$ point as the
vortex-antivortex pair.

As we shall show in section \ref{s4}, increasing the $x^1$ and $x^2$ radii
to arbitrary values $R_1$ and $R_2$ in the presence of $\alpha=1$
background can be described equivalently as increasing the radii of a pair
of different coordinates $\wh x^1$ and $\wh x^2$. But in these
coordinates, the original D2-$\bd$2 brane system, after being deformed by
the tachyon background \refb{e2.3} with $\alpha=1$, appears as a D0-$\bd$0
brane system situated at diametrically opposite points of the torus
spanned by $\wh x^1$ and $\wh x^2$. This shows that the spectrum of open
strings in the
background of a vortex-antivortex pair on the D2-$\bd$2 system wrapped
on a torus with radii $(R_1,R_2)$ is
identical to that on a D0-$\bd$0 brane system situated on a torus with the
same radii. This establishes the equivalence between the vortex-antivortex
pair on a D2-$\bd$2 brane system and D0-$\bd$0 brane pair.
\end{enumerate}

\sectiono{Conformal Field Theory at the Critical Radii} \label{s3}

In this section we shall study the marginal deformation of the BCFT on
the upper half plane at the
critical radii $R_1=R_2=1$ by
\refb{e2.3}. The relevant part of the BCFT at the critical radii before we
switch on the perturbation \refb{e2.3} is described by a pair of scalar
fields $X^i\equiv X^i_L+X^i_R$ for $i=1,2$ and their right- and
left-moving fermionic superpartners $\chi^i_R$, $\chi^i_L$. The Neumann 
boundary
condition satisfied by these fields on the real line is given by
\be \label{e3.1}
X^i_L=X^i_R\equiv {1\over 2} X^i_B, \qquad \chi^i_L=\chi^i_R\equiv
\chi^i_B\, .
\ee
We are considering here the NS
sector open string states. In the Ramond (R) sector we have a different
boundary condition on $\chi^i$, but it can be handled in a manner similar
to the one discussed in ref.\cite{9808141} and will not be discussed here.

Besides these fields we also have a time coordinate and its superpartners
satisfying Neumann boundary condition and 7 other space-like coordinates
and their superpartners satisfying Dirichlet boundary condition. We shall
refer to these fields as spectator fields as they do not play a major role
in the dynamics of the problem. We also have fermionic ghost fields
$b_L,c_L, b_R, c_R$ and bosonic ghost fields
$\beta_L,\gamma_L,\beta_R,\gamma_R$ satisfying Neumann boundary
condition.
We shall denote by
$\Phi_L$, $\Phi_R$ the left- and right-moving bosonized ghost
field of the $\beta,\gamma$ system\cite{FMS}, satisfying the
boundary condition
\be \label{eghb}
\Phi_L=\Phi_R\equiv \Phi_B\, .
\ee

The tachyon vertex operator in the $(-1)$ picture\cite{FMS} corresponding
to the deformation \refb{e2.3}
is given by:
\ben \label{e3.2}
V^{(-1)}_T &\propto& e^{-\Phi_B} \Big[ (e^{{i\over 2} (X^1_B+X^2_B)} -
e^{-{i\over 2}
(X^1_B+X^2_B)}) \otimes\sigma_1 \nonumber \\
&& + (e^{{i\over 2} (X^1_B-X^2_B)} - e^{-{i\over 2}
(X^1_B-X^2_B)}) \otimes\sigma_2 \Big]\, .
\een
In `zero' picture this vertex operator takes the form:
\ben \label{e3.3}
V^{(0)}_T &\propto& \Big[(\chi^1_B + \chi^2_B) 
(e^{{i\over 2} (X^1_B+X^2_B)} + e^{-{i\over 2}
(X^1_B+X^2_B)}) \otimes\sigma_1 \nonumber \\
&& +
(\chi^1_B - \chi^2_B) 
(e^{{i\over 2} (X^1_B-X^2_B)} + e^{-{i\over 2}
(X^1_B-X^2_B)}) \otimes\sigma_2\Big]\, .
\een

Let us now define a new set of variables:
\ben \label{e3.4}
Y^1 &=& {1\over \sqrt 2} (X^1 + X^2) \nonumber \\
Y^2 &=& {1\over \sqrt 2} (X^1 - X^2) \nonumber \\
\psi^1_R &=& {1\over \sqrt 2} (\chi^1_R + \chi^2_R) \nonumber \\
\psi^2_R &=& {1\over \sqrt 2} (\chi^1_R - \chi^2_R) \nonumber \\
\psi^1_L &=& {1\over \sqrt 2} (\chi^1_L + \chi^2_L) \nonumber \\
\psi^2_L &=& {1\over \sqrt 2} (\chi^1_L - \chi^2_L) \, .
\een
These fields satisfy the boundary conditions:
\be \label{e3.5}
Y^i_L=Y^i_R\equiv {1\over 2}Y^i_B, \qquad \psi^i_L=\psi^i_R\equiv
\psi^i_B\, , \qquad (i=1,2)
\ee
on the real line. In terms of these fields,
\be \label{e3.6}
V^{(0)}_T \propto \Big[ \psi^1_B(e^{{i\over\sqrt 2} Y^1_B} +
e^{-{i\over\sqrt 2}
Y^1_B})\otimes\sigma_1 +
\psi^2_B(e^{{i\over\sqrt 2} Y^2_B} + e^{-{i\over\sqrt 2}
Y^2_B})\otimes\sigma_2\Big]\, .
\ee 

We now fermionize the scalar fields $Y^i$ as follows:
\be \label{e16}
e^{i\sqrt 2 Y^i_R} = {1\over \sqrt 2} (\xi^i_R + i\eta^i_R)\otimes
\tau_i\, ,
\qquad
e^{i\sqrt 2 Y^i_L} = {1\over \sqrt 2} (\xi^i_L + i\eta^i_L) \otimes
\tau_i\, ,
\ee
where $\xi^i_R,\eta^i_R$ ($\xi^i_L$, $\eta^i_L$) are right- (left-)
moving
Majorana-Weyl fermions, and the Pauli matrices $\tau_i$ denote cocycle
factors\cite{COCYCLE} which must be put in to guarantee correct
(anti-)commutation
relations between various fields.\footnote{As we shall be using these
bosonization formulae to manipulate vertex operators in the BCFT
on the upper half plane, we only need to require that correct
(anti-)commutation relations are satisfied by the vertex operators
subject to the boundary condition \refb{e3.5}.} We also attach a cocycle
factor $\tau_3$ to $\psi^i_{L,R}$ and all the spectator fermions.
This guarantees
for example that $\psi^i$ and the spectator fermions commute
with both sides of eq.\refb{e16}. The cocycle factors should be taken to
commute with CP factors.

We can find another representation of the same conformal field theory by
rebosonising the fermions as follows:
\be \label{e22}
{1\over \sqrt 2} (\xi^i_R + i\psi^i_R) = e^{{i \sqrt 2}
\phi^i_R}\otimes \wt\tau_i\, ,
\qquad
{1\over \sqrt 2} (\xi^i_L + i\psi^i_L) = e^{{i \sqrt 2}
\phi^i_L}\otimes \wt\tau_i\, .
\ee
$\phi^1$, $\phi^2$ represent a pair of free scalar fields, and  
$\wt\tau_i$ are a new set of cocycle factors.
In this convention $\eta^1_{L,R}$, $\eta^2_{L,R}$
and spectator fermions carry the new cocycle factor
$\wt\tau_3$. 
There
is
a third representation in which we use a slightly different rebosonization:
\be \label{e22a}
{1\over \sqrt 2} (\eta^i_R + i\psi^i_R) = e^{{i \sqrt 2}
\phi^{i\prime}_R}\otimes \wh \tau_i\, ,
\qquad
{1\over \sqrt 2} (\eta^i_L + i\psi^i_L) = e^{{i \sqrt 2}
\phi^{i\prime}_L}\otimes \wh\tau_i\, ,
\ee
where $\phi^{i\prime}$ is another pair of scalar fields, and $\wh\tau_i$
denote another set of cocycle factors. $\xi^1_{L,R}$, $\xi^2_{L,R}$ and
the
spectator fermions will carry the cocycle factor $\wh\tau_3$ when we use
the set of variables $\phi^{i\prime}$, $\xi^i$ and the
spectator fields to describe the BCFT. We also define
\be \label{ephi}
\phi^i = \phi^i_L+\phi^i_R, \qquad \pip = \pip_L + \pip_R\, .
\ee

For later
use we list here the operator product expansions, and the relations
between the currents of free fermions and bosons:
\be \label{eyy1}
\psi^i_R(z)\psi^j_R(w)\simeq
\xi^i_R(z)\xi^j_R(w)\simeq\eta^i_R(z)\eta^j_R(w)\simeq
{i\over z-w}\delta_{ij} \, ,
\ee
\be \label{eyy2}
\p Y^i_R(z) \p Y^j_R(w) \simeq \p \phi^i_R(z) \p\phi^j_R(w) \simeq
\p\phi^{i\prime}_R(z)\p\phi^{j\prime}_R(w) \simeq -{1\over
2(z-w)^2}\delta_{ij}\, ,
\ee
\be \label{eyy3}
\psi^i_R\xi^i_R = i\sqrt 2 \p \phi^i_R\, , \qquad \eta^i_R\xi^i_R = i\sqrt
2 \p
Y^i_R\, ,
\qquad \psi^i_R\eta^i_R = i\sqrt 2 \p \phi^{i\prime}_R\, ,
\ee
with no summation over $i$ in the last equation.
Here $\simeq$ denotes equality up to non-singular terms. There are also
similar relations involving the left-moving currents.

Using eq.\refb{e16}
the boundary condition \refb{e3.5} on $Y^i$ can be translated to the
following boundary condition on the
fermions:
\be \label{e17}
\xi^i_L=\xi^i_R\equiv \xi^i_B\, , \qquad \eta^i_L=\eta^i_R 
\equiv \eta^i_B\, .
\ee
Combining these with
the boundary condition \refb{e3.5} on $\psi^i$,
we see from \refb{e22},
\refb{e22a} that
$\phi^i$ and $\pip$ both satisfy
Neumann boundary condition on the real line
\be \label{enn1}
\phi^i_L=\phi^{i}_R \equiv \phi^{i}_B / 2, \qquad
\phi^{i\prime}_L=\phi^{i\prime}_R \equiv \phi^{i\prime}_B / 2\, .
\ee

Using the bosonization relations \refb{e16}-\refb{e22a}, \refb{eyy3}, the
vertex operator $V^{(0)}_T$ given
in eq.\refb{e3.6} can be expressed as
\be \label{e21}
V^{(0)}_T \propto
[\psi^1_B\xi^1_B \otimes \sigma_1\otimes \tau_2 - \psi^2_B \xi^2_B \otimes
\sigma_2\otimes \tau_1]\, 
\propto
[\p \phi^1_B \otimes \sigma_1\otimes \tau_2 - \p\phi^2_B\otimes
\sigma_2\otimes\tau_1]\, .
\ee
$\partial$ denotes tangential
derivative along the boundary. The two operators appearing in
the right hand side of eq.\refb{e21} correspond to vertex operators of
constant U(1) gauge field
along
$\phi^1$ and $\phi^2$ directions, with generators $\sigma_1\times\tau_2$
and $\sigma_2\times\tau_1$ respectively. Since these two matrices commute,
we see
that switching on the tachyon vev corresponds to switching on a
pair of commuting
U(1) Wilson lines $A_{\phi_1}$ and $A_{\phi_2}$ along the bosonic
directions $\phi^1$ and $\phi^2$. This represents an exactly
marginal deformation. Switching on finite vacuum expectation
value (vev) of the tachyon then corresponds to inserting the operator
\be \label{e3.8}
\exp({i\alpha\over 2\sqrt 2}\ointop \p \phi^1_B \otimes \sigma_1\otimes
\tau_2 -
{i\alpha\over 2\sqrt 2} \ointop \p\phi^2_B\otimes
\sigma_2\otimes\tau_1)\, ,
\ee
where $\ointop$ denotes integration along the boundary.
Note that we have fixed the normalization of $\alpha$ in a specific
manner. This is the same normalization convention as in
ref.\cite{9808141}, and so we shall be able to use the results of
ref.\cite{9808141} directly. With this normalization, $\alpha$ is a
periodic variable with period 2.

Before we study the effect of switching
on such a tachyon vev on the spectrum of open strings, let us
study the open string spectrum
before switching on the tachyon vev.
Let us denote by $\HH$ the Hilbert space of states\footnote{Here we are
discussing open string states; so all states are created from $|0\rangle$
by vertex operators inserted at the boundary of the upper half plane.}
created by the half-integer moded $\psi^i$, $\xi^i$, $\eta^i$ oscillators
and the spectator fields on the $SL(2,R)$ invariant vacuum $|0\rangle$.
Alternatively we can also view $\HH$ as the Hilbert space of states
created from $|0\rangle$ by the spectator fields, together with vertex
operators involving $\psi^i$ and $Y^i$
with $Y^i$ momenta
quantized in units of
$1/\sqrt{2}$, or vertex operators involving $\eta^i$ and $\phi^i$
with ${\phi^i}$ momenta
quantized in units of $1/\sqrt{2}$, or vertex operators involving $\xi^i$
and $\pip$ with $\pip$ momenta quantized in units of $1/\sqrt{2}$. 
On $\HH$, we denote by $(-1)^F$ the world-sheet fermion
number
which changes the sign of $\psi^1$, $\psi^2$, and
the spectator fermions, leaving $\xi^i$, $\eta^i$ (and hence $Y^i$) and
the spectator bosons unchanged.
The SL(2,R) invariant
ground state $|0\rangle$ is taken to be odd under $(-1)^F$. 
We also define the transformations $h_1$ and $h_2$ as follows. Both $h_1$
and $h_2$ leave $\psi^1$, $\psi^2$ and all the spectator fermions
unchanged, but $h_1$ changes the sign of $\xi^1,\eta^1$ and $h_2$ changes
the sign of $\xi^2,\eta^2$. In terms of the bosonic variables $X^i$
or $Y^i$,
\ben \label{e3.7}
h_1 &:& X^1_{L,R}\to X^1_{L,R}+{\pi\over 2}, \quad X^2_{L,R}\to
X^2_{L,R}+ {\pi\over 2}, \nonumber \\ &&  Y^1_{L,R}\to
Y^1_{L,R}+ {\pi\over \sqrt{2}}, \quad Y^2_{L,R}\to Y^2_{L,R} \nonumber \\
h_2 &:& X^1_{L,R}\to X^1_{L,R}+{\pi\over 2}, \quad X^2_{L,R}\to
X^2_{L,R}-{\pi\over 2}, \nonumber \\ && Y^1_{L,R}\to
Y^1_{L,R}, \quad Y^2_{L,R}\to Y^2_{L,R} + {\pi\over\sqrt{2}}\, . \nonumber
\\
\een
{}It is easy to verify that with this definition, $(h_1)^2$ and
$(h_2)^2$ act as identity on all states. States with CP factor $I$
and $\sigma_3$ are $h_1h_2$ even and $(-1)^F$ even, whereas
states with CP factors $\sigma_1$ and $\sigma_2$ are $h_1h_2$ odd and
$(-1)^F$ odd. Taking into account the assignment
of the cocycle
factors ($\tau_1$ for $h_1$ odd states, $\tau_2$ for $h_2$ odd states and
$\tau_3$ for
$(-1)^F$ odd states) the quantum numbers carried by various open string
states, when expressed in terms of $\psi^i,\eta^i,\xi^i$ and the spectator
fields, are as given in table 1.\footnote{There will be extra cocycle
factors involving $\wt\tau_i$ when we express the vertex
operators in terms of the fields $\phi^i$, $\eta^i$ and
the spectator fields. But since these cocycle factors commute with those
carried by the gauge fields $A_{\phi^1}$ and $A_{\phi^2}$, we can ignore
them for the purpose of studying how the spectrum changes when we switch
on the gauge fields.} The complete spectrum of open string
states carrying a given CP$\otimes$cocycle factor is obtained by keeping
all states in $\HH$ carrying the quantum numbers
mentioned in the table. The last two columns describe if the open string
state is charged or neutral under the gauge fields $A_{\phi^1}$ or
$A_{\phi^2}$. This is easily determined by examining if the cocycles
factors commute with $\sigma_1\otimes\tau_2$ and $\sigma_2\otimes \tau_1$,
since these represent the U(1) generators associated with the background
gauge fields.

\begin{table}
\begin{eqnarray*} 
\begin{array}{||c|l|c|c|c|c||} \hline
  \rule{0mm}{5mm} {\hbox{CP}\otimes\hbox{cocycle}} & {(-1)^F}
  & {h_1} & {h_2} & A_{\phi^1} & A_{\phi^2}
  \\ \hline 
  \rule{0mm}{6mm} I\otimes I & \, \, \, \, \, + 
 & + &  + & n & n \\ \hline
  \rule{0mm}{6mm} I\otimes \tau_3 & \, \, \, \, \, + 
 & - &  - & c & c \\ \hline
  \rule{0mm}{6mm} \sigma_3 \otimes I & \, \, \, \, \, + 
 & + &  + & c & c \\ \hline
  \rule{0mm}{6mm} \sigma_3 \otimes \tau_3 & \, \, \, \, \, + 
 & - &  - & n & n \\ \hline
  \rule{0mm}{6mm} \sigma_1\otimes \tau_1 & \, \, \, \, \, - 
 & + &  - & c & c \\ \hline
  \rule{0mm}{6mm} \sigma_1\otimes \tau_2 & \, \, \, \, \, - 
 & - &  + & n & n \\ \hline
  \rule{0mm}{6mm} \sigma_2\otimes \tau_1 & \, \, \, \, \, - 
 & + &  - & n & n \\ \hline
  \rule{0mm}{6mm} \sigma_2\otimes \tau_2 & \, \, \, \, \, - 
 & - &  + & c & c \\ \hline
\end{array}
\end{eqnarray*} 
\caption{Quantum numbers carried by open string states with various CP
and cocycle factors. The last two columns indicate whether the
open string states are neutral ($n$) or charged ($c$) under the gauge
fields $A_{\phi^1}$ and $A_{\phi^2}$. Note that not all combinations of
CP$\otimes$cocycle factors are present in the spectrum.} 
\end{table}

{}From table 1 we note that before we switch on the background
\refb{e3.8}, each state is invariant under $(-1)^Fh_1h_2$, and that all
combinations of the quantum numbers subject to this condition appear
exactly twice in this
table. Thus the combined spectrum from all the sectors contain two copies
of $\HH$ with $(-1)^Fh_1h_2$ projection.
We also note that
open string states carrying CP$\otimes$cocycle
factors $I\otimes I$, $\sigma_3\otimes\tau_3$, $\sigma_1\otimes\tau_2$ and
$\sigma_2\otimes\tau_1$ are all neutral under both gauge fields. 
Combining the spectrum from all four neutral
sectors we see that each combination of quantum numbers (subject to the
condition $(-1)^Fh_1h_2=+1$) appears
exactly once; thus the combined spectrum of charge neutral states contain
a single copy of $\HH$ with $(-1)^F h_1 h_2$
projection. The combined spectrum of the charged states also contains
a single copy of $\HH$ with $(-1)^F h_1 h_2$
projection.

Let us now define
a new set of symmetry
generators $(-1)^{F_\phi}$, $h_{\phi^1}$ and $h_{\phi^2}$ on $\HH$ by
using the representation where $\HH$ is generated by
action on $|0\rangle$ by the vertex operator involving $\phi^i$, $\eta^i$
and the spectator
fields.
$(-1)^{F_\phi}$ changes the signs of $\eta^i$ and the spectator fermions,
leaving unchanged $\phi^i$ and the spectator bosons, 
and has eigenvalue $-1$ acting on the SL(2,R) invariant vacuum
$|0\rangle$.
$h_{\phi^i}$ leaves unchanged $\eta^1$, $\eta^2$ and all the spectator
fields, and transform $\phi^j$ as follows:
\ben \label{e3.9}
h_{\phi^1}  &:& \phi^1_{L,R} \to \phi^1_{L,R} + {\pi\over\sqrt{2}}\, ,
\quad \phi^2_{L,R}\to
\phi^2_{L,R}\, , \nonumber \\
h_{\phi^2}  &:& \phi^2_{L,R} \to \phi^2_{L,R} + {\pi\over\sqrt{2}}\, ,
\quad \phi^1_{L,R}\to
\phi^1_{L,R}\, . 
\een
Thus
$h_{\phi^i}$ changes the sign of $\xi^i$ and $\psi^i$.
With this definition, and using the bosonization relations
\refb{e16}, \refb{e22}, it is easy to
see that
\be \label{e3.10}
(-1)^F h_1 h_2 = (-1)^{F_\phi} h_{\phi^1} h_{\phi^2}\, .
\ee
(Both sides change the sign of $\xi^{1,2}$, $\eta^{1,2}$ and $\psi^{1,2}$,
and the spectator fermions, leaving the spectator bosons unchanged.)
Thus the combined spectrum of the charge neutral states as well as the
combined spectrum of the charged
states may be identified as a copy of $\HH$, subject
to the $(-1)^{F_\phi}h_{\phi^1}h_{\phi^2}$ projection.

Upon switching on the gauge fields, the spectrum in the charge neutral
sector does not change, but the spectrum in the charged sector
changes since the quantization laws for the $\phi^i$ momenta $p_{\phi^i}$ 
change. By combining the fields into charge eigenstates we can follow the
change in $p_{\phi_1}$, $p_{\phi_2}$ as a function of $\alpha$ and thus
completely determine the spectrum at every value of
$\alpha$.\footnote{With the normalization of $\alpha$ chosen here,
$\alpha=2$ corresponds to a shift in $p_{\phi^i}$ by $\pm\sqrt 2$. This
corresponds to states carrying same $h_{\phi^1}$ and $h_{\phi^2}$ quantum
numbers.
Hence $\alpha$ and $\alpha+2$ describes the same BCFT.} However
as we shall see in the next section, once we deform the radius away from
$R_1=R_2=1$ point, $\alpha=0$ and $1$  are the only inequivalent points
which give conformally invariant theories. Since $\alpha=0$ represents the
trivial tachyon background, we shall be interested in the $\alpha=1$
point. The spectrum at $\alpha=1$ simplifies enormously if we notice that
this corresponds to shifting the $p_{\phi^1}$ and
$p_{\phi^2}$ quantization laws by $\pm{1\over \sqrt 2}$\cite{9808141}, so
that its effect is to
simply reverse the sign of the $h_{\phi^1}$ and $h_{\phi^2}$ quantum
numbers. But the initial spectrum did not contain $h_{\phi^1}$ and
$h_{\phi^2}$ projections individually. It only contained
$(-1)^{F_\phi}h_{\phi^1}h_{\phi^2}$ projection, and this does not change.
As a result the spectrum at $\alpha=1$ is identical to the spectrum at
$\alpha=0$!

It may appear from this that at the end of the deformation the system has
come back to the original system! However, as we shall see in section
\ref{s4}, the response of this system to a change in the radii $R_1$ and
$R_2$ is very different from that of the original system. In order to
facilitate the analysis of section 4, we introduce dual coordinates
\be \label{e4.9}
\wh X^i_L = X^i_L, \qquad \wh X^i_R = - X^i_R\, .
\ee
In terms of the new coordinates $\wh X^i$, the BCFT at $\alpha=1$
corresponds to a
D0-brane
$\bd$0-brane pair, situated at diametrically opposite points of a square
torus with unit radii.

\sectiono{Deforming Away from the Critical Radius} \label{s4}

In this section we shall consider the effect of switching on the
perturbation that deforms the radii away from their critical values.
The procedure followed here will be similar to the one used in
ref.\cite{9808141}, so our discussion will be brief. There are four
possible marginal deformations of the bulk conformal field theory, three
of which correspond to deformation of the shape and size of the torus
labelled by the $X^1$-$X^2$ coordinates, and
the fourth one corresponds to switching on the anti-symmetric tensor field
in the $X^1$-$X^2$ plane. Using eqs.\refb{e3.4} we can express a general
perturbation of this kind in the (0,0) picture as
\be \label{e4.1}
K_{ij} \p X^i_L \p X^j_R = a_{ij} \p Y^i_L \p Y^j_R\, ,
\ee
where 
\be \label{e4.1a}
a = \pmatrix{{1\over \sqrt 2} & {1\over \sqrt 2} \cr
{1\over \sqrt 2} & -{1\over \sqrt 2}} K
\pmatrix{{1\over \sqrt 2} & {1\over \sqrt 2} \cr
{1\over \sqrt 2} & -{1\over \sqrt 2}}\, .
\ee

First we shall consider the effect of first order perturbation, and show
that in the presence of this perturbation the tachyon vertex operator
$V_T$ develops a one point function unless $\alpha=0$ or $\alpha=1$. The
procedure for doing this is similar to that discussed in
ref.\cite{9808141}. We insert a tachyon vertex operator $V^{(0)}_T$
given in eq.\refb{e21} at a point on the boundary of the disk (or upper
half plane), the
background \refb{e3.8} at the boundary of the disk, and a  
closed string vertex operator
corresponding to the perturbation \refb{e4.1} in the $(-1,-1)$ picture at
the center of the
disk. This is proportional to
\be \label{e4.2}
a_{ij} e^{-\Phi_L} e^{-\Phi_R} \psi^i_L \psi^j_R\, ,
\ee
where $\Phi_L$ and $\Phi_R$ are the left and right-moving bosonized ghost
fields\cite{FMS}. Computation of this amplitude is straightforward using
the description of the BCFT in terms of the $(\eta^i, \phi^i)$
fields\cite{9808141}. The $\phi^i$ momentum conservation laws tell us that
only the $i=j$ terms in \refb{e4.2} contribute.
The answer is proportional to
\be \label{e4.3}
(a_{11}+a_{22}) \sin(\pi\alpha)\, .
\ee
This shows that this one point function vanishes only for $\alpha=0$ and
$1$ (mod 2). It is also easy to check that at these two values of $\alpha$ 
the one
point function of all other open string vertex operators also vanish, and
hence these configurations describe consistent boundary conformal field
theories.

Next we need to study what happens when we go beyond the lowest order
perturbation in the deformation parameters $a_{ij}$. For this let us
consider an amplitude with an arbitrary number of insertions of the
operator \refb{e4.1} at various points in the interior of the disk,
insertion of the background 
\refb{e3.8} with $\alpha=1$ at the boundary of the disk, and a set of open
string vertex operators, whose correlation function we want to calculate,
at the boundary of the disk. We take two of these open string vertex
operators to be in the $-1$ picture and the rest in the zero picture
so that the total picture number of all the vertex operators add up
to $-2$. Again the calculation proceeds as
in ref.\cite{9808141}. We express all the operators in terms of $\eta^i$,
$\phi^i$, and the spectator fields. The vertex operator \refb{e4.1} can be
expressed as
products of $\eta^i_L$, $\eta^i_R$, $e^{\pm i\sqrt{2} \phi^i_L}$ and 
$e^{\pm i\sqrt{2} \phi^i_R}$ fields. The integrated vertex operator
\refb{e3.8} inserted at the boundary has two effects. It
effectively shifts the $\phi^i$ momenta of the open string vertex
operators as discussed in section \ref{s3}, and $\ointop\p \phi^i_B$ picks
up the $\phi^i$ winding number of all the closed string vertex operators
inserted in the interior of the disk. Using the operator product expansion
\refb{eyy2} it is easy to verify that ${1\over 2\sqrt{2}} \ointop \p
\phi^i_B={1\over \sqrt 2} \ointop \p \phi^i_R$, acting on any combination
of closed 
string vertex operators, gives an integral
multiple of $\pi$. Now, since
\be \label{e4.4} 
\exp(i \pi n \sigma_1\otimes\tau_2) = \exp(i \pi n),
\qquad 
\exp(i \pi n \sigma_2\otimes\tau_1) = \exp(i \pi n),
\quad \hbox{for integer} \, n\ee
we see that in taking into account the effect of \refb{e3.8} for
$\alpha=1$ on the closed
string vertex operators we can replace it by
\be \label{e4.5}
\exp({i\over 2\sqrt 2}\ointop \p \phi^1_B -
{i\over 2\sqrt 2} \ointop \p\phi^2_B) =
\exp({i\over \sqrt 2}\ointop \p \phi^1_R -
{i\over \sqrt 2} \ointop \p\phi^2_R)\, .
\ee
In going from the left hand side to the right hand side of eq.\refb{e4.5}
we have used the boundary condition \refb{enn1}. Since $\phi^i_R$ are
holomorphic fields,
we can now deform the integration contour into the interior of the disk,
picking up residues from the location of the closed string vertex
operators. The net effect is to replace each insertion of \refb{e4.1} by
\be \label{e4.6}
\exp\Big({i\over \sqrt 2}\ointop \p \phi^1_R -
{i\over \sqrt 2} \ointop \p\phi^2_R\Big)
a_{ij} \p Y^i_L \p Y^j_R\, ,
\ee
where the contours of integration in the exponent are around the location
of the closed string vertex operator. This can be easily evaluated using
eqs.\refb{e22}-\refb{eyy3}, and
the result is\footnote{Here we have differed somewhat from the analysis of
ref.\cite{9808141}. In \cite{9808141} $\ointop \p\phi_B$ in the exponent
was replaced by $\ointop(\p\phi_L+\p\phi_R)$ instead of
$2\ointop\p\phi_R$. If we follow this procedure here, then eq.\refb{e4.7}
will be replaced by $-a_{ij}\p\pip_L\p\phi^{j\prime}_R$. This would be
analogous to the corresponding result in \cite{9808141}. The final result
for the spectrum and correlation functions however does not depend on
which procedure we follow, since the two procedures are related by the
symmetry transformation $(\phi^i_L,\phi^i_R)\to (\phi^i_L +{\pi\over
2\sqrt 2}, \phi^i_R + {\pi\over 2 \sqrt 2})$.} \be \label{e4.7}
- a_{ij} \p Y^i_L \p Y^j_R =
- K_{ij} \p  X^i_L \p X^j_R \, .
\ee
In terms of the coordinates $\wh X^i$ introduced in eq.\refb{e4.9},
the right hand side of eq.\refb{e4.7} can be expressed as
\be \label{e4.8aa}
K_{ij} \p \wh X^i_L \p \wh X^j_R \, .
\ee
This has the same form as the left hand side of \refb{e4.1} with $X^i$
replaced by $\wh X^i$.  
Thus deforming the $X^i$ radius from 1 to $R_i$ in
the
presence of the tachyon background \refb{e3.8} with $\alpha=1$ corresponds
to deforming
the $\wh X^i$ radius from 1 to $R_i$. The D-brane system now describes a
D0-brane $\bd$0-brane pair situated at diametrically opposite points of
this torus. 

This shows that the spectrum of open string states living on a vortex
antivortex pair at the diametrically opposite points on a D2-$\bd$2 system
wrapped on a torus is identical to the spectrum of open string states
living on a D0-$\bd$0 brane pair at the diametrically opposite points of
the same torus. This leads to the identification of the vortex
(antivortex) on a D2-$\bd$2 system with a D0 ($\bd$0) brane.

\sectiono{Generalizations} \label{s5}

The method used here can easily be generalized to prove the identification
of a vortex solution on a D$p$-$\bd p$ brane pair with a D-$(p-2)$ brane.
The analysis is exactly identical; all that is required is to change the
Dirichlet boundary condition on $(p-2)$ of the spectator superfields to
Neumann boundary condition.

It is also possible to generalize this
analysis to show that a codimention $2n$
soliton on $2^{n-1}$ pairs of D$p$-$\bd p$ brane system represents a
D$(p-2n)$ brane\cite{9808141,9810188}. Again for simplicity let us assume
that all the spectator fields except the time direction has Dirichlet
boundary condition, {\it i..e.} take $p=2n$. We compactify each of the
$2n$ directions tangential to the D-branes on a circle of radius 1. Let us
label these coordinates by $x^1,\ldots x^{2n}$. The open string states
carry $2^n\times 2^n$ CP factors; with the tachyon states carrying off
diagonal CP factors of the form:
\be \label{e5.1}
\pmatrix{0 & A\cr A^\dagger & 0}\, ,
\ee
where $A$ is a $2^{n-1}\times 2^{n-1}$ complex matrix. Under the
$U(2^{n-1})\times U(2^{n-1})$
gauge transformation on the brane-antibrane system, generated by
$U(2^{n-1})$
matrices $U$ and $V$,
\be \label{e5.2}
A \to U A V^\dagger\, .
\ee
Let us now, following ref.\cite{9810188}, choose a
$2^n\times 2^n$ dimensional representation of the
$2n$ dimensional Clifford algebra in which each of the $\Gamma$-matrices
has the form given in \refb{e5.1}. Let $X^i$ ($1\le i\le 2n$) denote the
coordinate fields tangential to the D-brane, and $\chi^i_R$, $\chi^i_L$
be
their right and left-moving superpartners. Let us define
\ben \label{e5.3}
&& Y^{2k+1}= {1\over \sqrt 2} (X^{2k+1}+ X^{2k+2})\, , \qquad
Y^{2k+2}= {1\over \sqrt 2} (X^{2k+1} - X^{2k+2})\, , \nonumber \\
&& \psi^{2k+1}_R= {1\over \sqrt 2} (\chi^{2k+1}_R+ \chi^{2k+2}_R)\, ,
\qquad
\psi^{2k+2}_R= {1\over \sqrt 2} (\chi^{2k+1}_R - \chi^{2k+2}_R)\, ,
\nonumber \\
&& \psi^{2k+1}_L= {1\over \sqrt 2} (\chi^{2k+1}_L+ \chi^{2k+2}_L)\, ,
\qquad
\psi^{2k+2}_L= {1\over \sqrt 2} (\chi^{2k+1}_L -\chi^{2k+2}_L)\, ,
\nonumber \\
\een
for $0\le k\le (n-1)$.
Consider now the following vertex operator in the $(-1)$ picture
\be \label{e5.4}
V_T^{(-1)} \propto e^{-\Phi_B}\, \sum_{i=1}^{2n} \sin{Y_B^i\over \sqrt 2}
\otimes
\Gamma_i\, ,
\ee
where the subscript $B$ denotes the boundary value as usual. In order that
this is an allowed vertex operator, we need to switch on appropriate
Wilson lines on the branes; we assume that this has been done. (We
can, for example, take the Wilson line along the $X^{2k+1}$  and the
$X^{2k+2}$ direction to be $\Gamma^{2k+1}\Gamma^{2k+2}$.
Conjugation by this matrix changes the sign of $\Gamma^{2k+1}$ and
$\Gamma^{2k+2}$, keeping the other $\Gamma^i$'s invariant. Unlike in the
case analysed earlier, this amounts to switching on Wilson lines both on
the D-branes and the anti-D-branes.)  In the zero picture the vertex
operator \refb{e5.4} takes the form: \be \label{e5.5}
V_T^{(0)} \propto \sum_{i=1}^{2n} \psi^i_B \cos{Y_B^i\over \sqrt 2}
\otimes
\Gamma_i\, .
\ee
It is easy to see via the bosonization procedure that each term in the sum
represents switching on a constant gauge field (Wilson line) along a new
bosonic coordinate and
hence is a marginal deformation.
Furthermore, since all the $\Gamma$
matrices anti-commute with each other, and the fermions $\psi^i_B$ also
anti-commute with each other, we see that the different terms in the sum
commute with each other. After bosonization this will imply that the
CP$\otimes$cocycle factors carried by the different gauge fields commute
with each other. Thus \refb{e5.5} corresponds to switching on constant,
commuting gauge fields along different directions, and represents a
marginal deformation.

In analogy with the analysis of sections \ref{s3}, \ref{s4}, one expects
that the BCFT at the end of the marginal deformation (the $\alpha=1$
point) is more naturally described in terms of a set of dual coordinates
$\wh x^i$
in which the system appears
as $2^{n-1}$ D0-$\bd$0 brane pair. If we
increase the radii of the original torus to arbitrary values $R_i$ after
the marginal deformation, then it would correspond to increasing the radii
of the torus described by the $\wh x^i$ coordinates to the same values
$R_i$ as in section \ref{s4}. Thus the end result will be $2^{n-1}$
D0-$\bd$0 brane pairs
situated at different point on the torus with radii $R_i$.\footnote{A  
consistency check for this
scenario is that at the critical radii the total 
mass of the initial
configuration is given by $2^n/g$, where $g$ denotes the closed string
coupling constant, since each of the $2^n$  
D-$2n$-branes / $\bd$-$2n$-branes
has area equal to $(2\pi)^{2n}$ and tension equal to $1/((2\pi)^{2n}g)$.
This agrees with the total mass of $2^{n-1}$ D0-$\bd$0 brane pair. Also
both the initial and the final state has vanishing RR charge.}

On the other hand,
we can examine the tachyon background \refb{e5.4} and try to identify
it as a collection of solitons. For this we need to identify the soliton
cores as the places where the tachyon field vanishes. This requires
each $y^i$ to be an integral multiple of $\pi\sqrt 2$. Using
eqs.\refb{e5.3}, and that each $x^i$ describes a circle of radius 1, we
see that the inequivalent points are given by all possible combination of
the configurations
\be \label{e5.6}
(x^{2k+1}, x^{2k+2}) = (0, 0) \quad \hbox{or} \quad (\pi, \pi)\, .
\ee
Since there are $n$ pairs of coordinates, 
this gives $2^n$ possibilities. To study the nature of the solution near
the core, let us consider one of them, {\it e.g.} $(x^1,\ldots x^{2n}) =
(0,
\ldots 0)$. The tachyon background near this point is proportional to
\be \label{e5.7}
y^i \Gamma_i\, .
\ee
This is precisely the form suggested in ref.\cite{9810188} for a
codimension $2n$ soliton on $2^{n-1}$ brane-antibrane pairs. 
On the other hand, expanding the tachyon field near the point
$(x^1,x^2\ldots x^{2n})=(\pi,\pi, 0, 0, \ldots 0)$, we get the tachyon
background to be proportional to:
\be \label{e5.8}
-(y^1-\pi\sqrt 2)\Gamma^1 + \sum_{i=2}^{2n} y^i \Gamma^i\, .
\ee
The $-$ sign in front of the first term indicates that it has opposite
orientation compared to \refb{e5.7} and represents an anti-soliton. In
general the solution around a point \refb{e5.6} represents a soliton
(anti-soliton) if the number of coordinate pairs taking the value
$(\pi,\pi)$ is even (odd). Thus the tachyon background \refb{e5.5}
represents creation of $2^{n-1}$ soliton - antisoliton pairs. {}From this 
we can
conclude that $2^{n-1}$ soliton - antisoliton pairs represent $2^{n-1}$
D0-$\bd$0 pairs, and hence a single D0-brane should be identified with the
single codimension $2n$ soliton on the D-$2n$-brane $\bd$-$2n$-brane pair. 

Generalization of this result to solitons of odd codimension, which are
expected to describe non-BPS D-branes\cite{9810188,9812135} is also
straightforward. In this case we take $2^{n-1}$ D-$(2n-1)$-brane
$\bd$-$(2n-1)$-brane pairs in type IIB string theory along $x^1,\ldots
x^{2n-1}$, with $x^1,\ldots x^{2n-2}$ directions compactified on circles
of radii 1 and $x^{2n-1}$ direction compactified on a circle of radius
${1\over \sqrt 2}$. We define $Y^i$'s for $1\le i\le (2n-2)$ as in
eq.\refb{e5.3} and consider tachyon background associated with the vertex
operator\footnote{This requires choosing the Wilson lines along
$x^1,\ldots x^{2n-1}$ as before. Since the Wilson line along $x^{2n-1}$
requires a $\Gamma_{2n}$ we see that we need a representation of the $2n$
dimensional Clifford algebra even though we have only $(2n-1)$
coordinates.}
\be \label{enb1}
V_T^{(-1)} \propto e^{-\Phi_B}\, \Big[ \sum_{i=1}^{2n-2} \sin{Y_B^i\over
\sqrt 2}
\otimes
\Gamma_i + \sin{X_B\over \sqrt 2} \otimes \Gamma_{2n-1}\Big]\, ,
\ee
This corresponds to creation of $2^{n-1}$ solitons at
\ben \label{enb2}
&& (x^{2k+1}, x^{2k+2}) = (0,0) \quad \hbox{or} \quad (\pi, \pi)
\quad \hbox{for} \quad 0\le k\le (n-2)\, ,\nonumber
\\
&& x^{2n-1} = 0\, .
\een
On the other hand it is easy to see that the vertex operator \refb{enb1},
in the zero picture, represents a marginal deformation. We expect this to
convert the original brane configuration to a configuration of non-BPS
branes as in ref.\cite{9808141}. The number of such D0-branes can easily
be seen to be $2^{n-1}$ by comparing the masses of the initial and the
final configurations. This shows the equivalence between the non-BPS
D0-brane and a soliton on $2^{n-1}$ D-$(2n-1)$ $-$ $\bd$-$(2n-1)$ brane
pairs.

\end{document}